\documentclass[12pt,preprint]{aastex}
\tighten
\eqsecnum
\received{}
\accepted{}
\journalid{}{}
\articleid{}{}
\shortauthors{ }
\shorttitle{UMa Moving Group Abundances} 

\begin{document}

\title{High-Resolution Spectroscopy of Ursa Major Moving Group Stars\altaffilmark{1,2}}

\author{Jeremy R. King \& Simon C. Schuler}
\affil{Department of Physics and Astronomy, 118 Kinard Laboratory,\\
Clemson University, Clemson, SC{\ \ }29634-0978}
\email{jking2,sschule@ces.clemson.edu}

\altaffiltext{1}{This paper includes data taken at The McDonald Observatory of 
The University of Texas at Austin.}
\altaffiltext{2}{Based on observations obtained at Kitt Peak National Observatory, 
a division of the National Optical Astronomy Observatories, which is operated by 
the Association of Universities for Research in Astronomy, Inc.~under cooperative 
agreement with the National Science Foundation.}

\begin{abstract}

We use new and extant literature spectroscopy to address abundances and membership for 
UMa moving group stars.  We first compare the UMa, Coma, and Hyades H-R diagrams via
a homogeneous set of isochrones, and find that these three aggregates are essentially 
coeval; this (near) coevality can explain the indistinguishable distributions 
of UMa and Hyades dwarfs in the chromospheric emission versus color plane.  Our 
spectroscopy of cool UMa dwarfs reveals striking abundance anomalies--trends 
with $T_{\rm eff}$, ionization state, and excitation potential--like those recently seen 
in young cool M34, Pleaides, and Hyades dwarfs.  In particular, the trend of rising  
${\lambda}7774$-based \ion{O}{1} abundance with declining $T_{\rm eff}$ is markedly 
subdued in UMa compared to the Pleiades, suggesting a dependence on age or metallicity. 
Recent photometric metallicity estimates for several UMa dwarfs are markedly low compared 
to the group's canonical metallicity, and similar deviants are seen among cool Hyads as 
well.  Our spectroscopy does not confirm these curious photometric estimates, which seem 
to be called into question for cool dwarfs.  Despite disparate sources of Li data, 
our homogeneous analysis indicates that UMa members evince remarkably small scatter 
in the Li-$T_{\rm eff}$ plane for $T_{\rm eff}{\ge}5200$ K.  Significant star-to-star 
scatter suggested by previous studies is seen for cooler stars.  Comparison with the  
consistently determined Hyades Li-$T_{\rm eff}$ trend reveals differences qualitatively 
consistent with this cluster's larger [Fe/H] (and perhaps slightly larger age).  However, 
quantitative comparison with standard stellar models indicates the differences are smaller 
than expected, suggesting the action of a fourth parameter beyond age, mass, and [Fe/H] 
controlling Li depletion.  The UMa-Coma cool star Li abundances may show a slight 0.2 dex 
difference; however, this may be mass-independent, and thus more consistent with a modest 
initial Li abundance difference. 
\end{abstract}


\section{Introduction}

The complex patterns exhibited by Li abundances in solar-type stars present an ongoing challenge to 
our fundamental understanding of stellar physics, spectral line formation, and Galactic chemical 
evolution.  Open clusters are important objects for deciphering these patterns since they provide a 
large number of stars with presumably identical heavy element composition, initial Li abundance, but 
differing mass; moreover, these objects can be accurately dated-- at least in a relative sense.  
Open clusters thus provide a unique and valuable means to study two critical problems related to 
stellar Li abundances:  a) the large star-to-star scatter seen in late-G and K dwarfs and connections 
to scatter in other stellar properties (Soderblom et al.~1993a; King, Krishnamurthi \& Pinsonneault 
2000), and b) deconvolving the effects of age-dependent main-sequence depletion and opacity-dependent 
pre-main-sequence depletion mechanisms in producing intercluster differences in the Li-mass profile 
(Soderblom 1993b, Swenson et al.~1994, Piau \& Turck-Chieze 2002).   

\citet{SPFJ} have noted the important niche in such attempts played by the UMa moving group: most 
notable is its ability to serve as a proxy for a cluster with an age presumably intermediate to the 
nearby and well-studied Pleaides and Hyades clusters, but having a subsolar ``metallicity'' 
($-0.08$; Boesgaard, Budge, \& Burck 1988; Boesgaard \& Friel 1990) lower than either cluster.  
Li abundances in the UMa moving group have been studied previously by \citet{BBB} and \citet{SPFJ}.  
The intervening decade following these studies has seen the availability of new UMa star data-- 
activity measures, radial velocities, photometry, Hipparcos parallaxes, etc-- which can be used to 
refine moving group membership. Here, we use new membership information, homogeneously-analyzed 
abundance data from the literature, and original spectroscopy of our own to revisit Li abundances 
in the UMa group. 

\section{Data and Analysis}

We pulled our UMa stellar sample from the recent membership study of \citet{King03}.  Stars with probable 
and possible final membership status (their `Y' and `Y?' classes) were selected, and Li measurements 
searched for in the literature.  Table 7 lists the full sample of stars considered here, their $B-V$ color 
(from the tabulation in \citet{King03}, projected rotational velocity, \ion{Ca}{2} chromospheric emission 
index, ${\lambda}6707$ \ion{Li}{1} equivalent widths, effective temperatures, LTE Li abundances, and 
associated references.  Several of the Li measurements are actually for the \ion{Li}{1} and neighboring 
${\lambda}6707.4$ \ion{Fe}{1} feature.  These cases were noted and corrected for in the analysis. 

Seven additional UMa candidates from \citet{King03} were selected for spectroscopic study utilizing new 
original data.  HD 28495, 59747, and 173950 were classified by \citet{King03} as members.  The stars 
HD 63433 and 75935 were deemed kinematic members, but photometric membership was ambiguous.  HD 81659 and 
167389 were considered kinematic non-members, but \citet{M01} classified them as kinematic members; our hope 
was to bring abundance data to bear on the issue of membership for the latter four stars.  These 7 additional
stars are listed at the bottom of Table 7.     

We obtained spectroscopy of 3 of the additional UMa candidates in October 2004 with the 
``2dcoude'' cross-dispersed echelle spectrometer on the 2.7m Harlan J. Smith Telescope at McDonald 
Observatory and a thinned Tektronix 2048${\times}$2048 CCD having 24 ${\mu}$m pixels.  Use of the folded 
Schmidt camera and chosen slit yielded a 2-pixel spectral resolution of $R{\sim}60,000$.  The resulting 
per pixel S/N values in the continuum regions near the 6707 {\AA} \ion{Li}{1} region were 200, 240, and 
310 for HD 28495, 167389, and 173950.  Spectroscopy of the other 4 additional UMa candidates was secured 
in December 2004 with the Cassegrain echelle spectrograph on the KPNO 4m.  The instrumental setup consisted 
of the 58-63 echelle grating, 226-1 cross disperser, long-focus camera, and T2KB Tektronix 2048${\times}$2048 
CCD; a 0.9 arcsec slit width yielded a spectral resolution of $R{\sim}40,000$.  The per pixel S/N 
values in the \ion{Li}{1} region were 465, 235, 270, and 155 for HD 59747, 63433, 75935, and 81659.  
Data reduction was carried out with standard routines in the {\sf IRAF} package.  Sample spectra are 
shown in Figure 1.  Equivalent widths were measured with the profile fitting routines in the 1-d spectral 
analysis package {\sf SPECTRE} \citep{FS}, and are listed in Tables 1, 3, 5, 6, and 7.    
\marginpar{Fig.~1}
\marginpar{Tab.~1}

Li abundances were determined from the equivalent widths using the {\sf{LIFIND}} software package 
kindly provided by Dr.~A.~Steinhauer (2003).  The program determines a color-based effective temperature \\ 
$$T_{\rm eff}=8344-3631.32{\times}(B-V)-2211.81{\times}(B-V)^2+3265.44{\times}(B-V)^3-1033.62{\times}(B-V)^4+701.7{\times}(B-V){\times}([Fe/H]-[Fe/H]_{\rm Hyades})$$
This relation is a slightly higher order fit to the same calibrating data used by \citet{DSJ}; the 
zero-point and metallicity terms are discussed by \cite{DKBR}.  Interpolating within an internal library 
of curves of growth generated by the LTE analysis package {\sf{MOOG}} using Kurucz (1992; private 
communication) model atmospheres, {\sf{LIFIND}} then returns a Li abundance for a given input equivalent 
width and $T_{\rm eff}$.  When required, {\sf{LIFIND}} also corrects the Li abundance for contributions 
in the ${\lambda}6707.4$ region typically dominated by an \ion{Fe}{1} feature; we assumed [m/H]$=-0.08$ 
for these corrections.  For stars with multiple Li measurements, we simply averaged the resulting abundances
together.  The Li abundance uncertainties listed in Table 7 are internal values comprising contributions due to 
internal $T_{\rm eff}$ uncertainties from photometric uncertainties and to equivalent width uncertainties.  
The latter were gauged from multiple measurements, taken from listed uncertainties in the original sources, 
or calculated from reported S/N, instrumental dispersion, and spectral resolution (or FWHM in the case of 
non-negligibly rotating stars) values via the formalism of \citet{Cay88}.  If one wishes to consider the 
absolute Li abundances alone or compare these to other analyses using different $T_{\rm eff}$ scales, then 
a larger total $T_{\rm eff}$ uncertainty of ${\sim}100$ K is more appropriate.  This increases the Li 
abundance uncertainties by ${\sim}0.05$ dex given the $T_{\rm eff}$ sensitivity of the derived Li 
abundance in our stars (a change of ${\pm}0.12$ dex and ${\pm}0.09$ dex in log $N$(Li) for a change of ${\pm}100$ K
in $T_{\rm eff}$ at 5100 and 5750 K, respectively).   

The derivation of O, Ca, Cr, Fe, and Ni abundances in our 7 additional UMa candidates proceeded as follows.  
$T_{\rm eff}$ values were taken from above and combined with log $g$ values from Yale-Yonsei isochrones 
(see next section).  Microturbulent velocities were then calculated from the relation of \citet{AP04}.  
These stellar parameters and the overall metallicity of the model atmosphere are listed in Table 2.  The
lines listed in Tables 1 and 3 are allegedly clean ``case a'' lines from \citet{Th90}, from which we 
also took oscillator strengths.  We carried out a differential analysis relative to Sun in order to 
minimize the effects of oscillator strength errors.  This was done by measuring the same lines in a spectrum 
of the zenith daytime sky obtained at the McDonald 2.7m during our October 2004 run, and analyzing them
in the same fashion.  Abundances were derived using the 2002 version of the LTE analysis package {\sf MOOG} 
and Kurucz (1992; private communication) model atmospheres.   Absolute solar abundances, log $N$(X), and
relative stellar abundances normalized to solar values on a line-by-line basis, [X/H], are given in 
Tables 2 and 4.  Table 5 contains our O results from the high excitation ${\lambda}7774$ triplet.  Table 6 contains
our results for \ion{Fe}{2}; while the ${\lambda}6416$ \ion{Fe}{2} feature appears clean and unblended in all of 
our spectra (and high resolution solar atlases), the behavior of its associated abundances relative to those from 
the other 3 lines may suggest mild contamination of the former by another low excitation transition.   

The neutral lines of Ca, Cr, Fe, and Ni demonstrate a derived abundance sensitivity of ${\pm}0.05$ and ${\pm}0.08$ dex
for a ${\pm}100$ K change in $T_{\rm eff}$ at 5100 and 5750 K, respectively.  The corresponding O sensitivities 
are ${\mp}0.13$ and ${\mp}0.09$ dex; those for \ion{Fe}{2} are ${\mp}0.10$ and ${\mp}0.04$ dex.  These sensitivities, 
the internal $T_{\rm eff}$ uncertainties of ${\sim}45$ K, and the small internal mean measurement uncertainties (typically
a couple hundredths of a dex), yield total internal uncertainties in [X/H] for all species in the 0.05-0.08 range.  As
for Li, these uncertainties are appropriate for examining star-to-star scatter.  For the purpose of external comparisons, 
total $T_{\rm eff}$ uncertaines of  ${\sim}100$ K are more appropriate.  These bring total abundance uncertainties to the 0.10-0.12 
dex level.   
\marginpar{Tab.~2}
\marginpar{Tab.~3}
\marginpar{Tab.~4}
\marginpar{Tab.~5}
\marginpar{Tab.~6}

\section{Results and Discussion}

\subsection{The Relative Age of UMa} 

Before discussing age-related implications of the UMa Li-$T_{\rm eff}$ morphology, it is 
useful to revisit the relative age of UMa and two key clusters-- the Hyades and Coma
Berenices.  The right hand panel of Figure 2 shows the color-magnitude 
diagram of the Hyades using the ``high fidelity'' sample and {\it Hipparcos\/} parallaxes from
\citet{dB01}.  The lines are the 500, 700, and 900 Myr, [Fe/H]$=+0.13$, [$\alpha$/Fe]$=0$ Yale-Yonsei 
isochrones \citep{YY} using the Lejeune et al.~(1998) color-$T_{\rm eff}$ relations.  The left hand
panel shows the photometry from the final {\it Hipparcos\/}-based UMa member sample of \citet{King03} 
and the [Fe/H]$=-0.08$, [$\alpha$/Fe]$=0$ Yale-Yonsei isochrones for 400, 600, and 800 Myr (all using 
the Lejeune et al.~1998 color-temperature relations). 
\marginpar{Fig.~2}

Earlier inhomogeneous age estimates placed the Hyades-UMa age difference at 300-500 Myr.  
The homogeneous comparison in Figure 2 suggests that this age difference is, in fact, 
considerably smaller-- ${\le}100$ Myr, but the uncertainties may even allow coevality.  Assuming 
a significant UMa-Hyades age difference, \citet{SC87} called attention to the seemingly remarkable 
similarity of the mean UMa and Hyades chromospheric emission levels.  Scatter in the UMa emission levels 
is also significantly smaller than in younger clusters such as the Pleiades (e.g., Figure 11 of King et 
al.~2003).  A similar age for UMa and the Hyades, suggested here, at last provides a natural explanation 
for these observations.  

The left hand panel of Figure 3 shows again the UMa color-magnitude diagram, while the right 
hand panel shows that for the Coma Berenices cluster.  The Coma photometry is from \citet{JK55} 
and \citet{FJJB}, and the assumed reddening of $E(B-V)=0$ and distance modulus of 4.54 are taken 
from \citet{Pin98}.  Given a Coma metal abundance of [Fe/H]$=-0.07$ \citep{B89}, we utilized the 
same isochrones displayed in the UMa diagram.  The Coma-UMa age comparison in Figure 3 suggests that 
these two systems too are essentially coeval.   
\marginpar{Fig.~3} 

\subsection{Metal Abundances and Membership}

Table 7 lists photometric and spectroscopic [Fe/H] values for our UMa stars.  The former are 
Stromgren-based values from the recent large survey of \citet{Nord04}.  The latter are taken
from literature values tabulated in \citet{K03} or, for the additional UMa stars at the bottom
of the table, our own results in Tables 2 and 4.  There are several stars for which the photometric 
metallicity is notably lower than the canonical spectroscopic value of [Fe/H]$=-0.09$ \citep{BF90}:  
HD 11131 (-0.27), 109799 (-0.24), 184960 (-0.32), 28495 (-0.41), and 173950 (-0.43).  As a check on 
these photometric metallicities, we calculated the mean abundance of Hyades members in the 
\citet{Nord04} catalog using the cluster membership list of \citet{Perry98} culled of questionable 
members. The result is [Fe/H]$=-0.01$ with a star-to-star scatter of 0.14 dex; this mean, which is 
raised by only 0.02 dex if spectroscopic binaries and radial velocity variables are excluded, is 
some ${\sim}0.15$ dex lower than the canonical spectroscopic Hyades metallicity.  It is not clear 
that the \citet{Nord04} abundance data are robust enough to address UMa membership. 
\marginpar{Tab.~7}

From our own spectroscopic abundance results in Tables 2 and 4, several things seem clear.  First, 
the line-to-line scatter in the abundances is satisfyingly small.  This suggests that insidious effects 
noted by King et al.~(2000), such as differential blending in the Sun relative to the cooler additional 
UMa stars, is not important here.  However, an example of the pitfalls awaiting the unwary spectroscopist is 
provided by the ${\lambda}6417$ \ion{Ca}{1} line, which is blended in our cool UMa candidates, but 
apparently clean in the Sun. The blend is subtle, particularly given finite S/N, but identified
from the consistent appearance of the line profile in all cool stars and consistently grossly deviant 
abundances.  It is possible that a few, even more subtle, ``clunkers'' have escaped detection and
reside in Table 1 and 3. We simply note again that differential blending is a potential pitfall 
in differential analyses of cool stars--particularly when using lines from a Sun-based line list.    

Second, both the photometric and our own spectroscopic abundances indicate that HD 81659 is markedly 
metal-rich compared to true UMa group stars. Given our earlier kinematic non-membership assignment, we
eliminate it as an UMa group member.  Indeed, its Li abundance is markedly lower than UMa stars
of similar $T_{\rm eff}$ (see below).

Third, the markedly low photometric abundances for HD 28495 and 173950 are not confirmed by
spectroscopic analysis.  Our [Fe/H] values are some ${\sim}0.25$ dex higher.  While considerably
more analysis with larger samples of stars is needed, this could signal a problem with the
photometric determinations of cool ($T_{\rm eff}{\le}5200$ K) stars; indeed, the markedly low 
photometric [Fe/H] values for Hyades members from \citet{Nord04} seem to occur preferentially at the cool end.  

Fourth, striking evidence for overexcitation/ionization is clearly seen in our cool UMa stars.  A 
growing body of work (Schuler et al.~2003, 2004; Morel \& Micela 2004; Morel et al.~2004; Yong et al.~2004) 
building on earlier suggestions (Cayrel et al.~1985, King et al.~2000) indicates that cool dwarf 
abundances show excitation and ionization-related anomalies when subjected to LTE analysis with 
standard stellar photospheric models.  This behavior is seen in our stars:  the Ca abundances derived 
from the higher excitation ${\lambda}6417$ feature (measured to account for the blending noted above) 
are consistently higher than derived from other Ca transitions for the cool stars in Table 4; the Cr 
abundancs derived from the lower excitation ${\lambda}6330$ feature are consistently lower than derived 
from other Cr transitions in all cases (Table 4).  The \ion{Fe}{1} abundances derived from the lower 
excitation ${\lambda}6498$ feature are lower than derived from higher excitation features in all cases 
(Table 4).  More marked is the \ion{Fe}{2}$-$\ion{Fe}{1} difference--a stunning 0.42 dex for the four 
coolest objects in Table 6.  

Figure 4 shows the \ion{Fe}{2}-\ion{Fe}{1} differences (see Table 6) and high excitation (9 eV) ${\lambda}7774$ 
triplet-based [O/H] values (see Table 5) versus $T_{\rm eff}$ for HD 28495, 59747, 63433, 75935, 167389, and 173950 
derived from our new spectroscopy.  There is a clear trend with $T{\rm eff}$, and that for \ion{O}{1} appears notably
more shallow than for the Pleiades.  These striking trends are qualitatively similar to those for 
[O/H] and \ion{Fe}{1}-\ion{Fe}{2} seen by Schuler et al.~(2003, 2004) and Yong et al.~(2004) for M34, the
Pleiades, and the Hyades.   Whether the shallower slope of the \ion{O}{1} trend for our UMa stars compared
to the Pleiades is somehow related to the former's larger age or lower metallicity is unclear, and 
will require observations of additional clusters/moving groups.   
\marginpar{Fig.~4}

We believe that the anomalous abundance results for cool stars like that shown in Figure 4 are not 
explained by simple modest parameter variations.  For example, the large \ion{O}{1} abundances and
\ion{Fe}{2}-\ion{Fe}{1} differences could be removed by lowering log $g$ by in the cool stars-- but
by a full dex.  Raising the overall $T_{\rm eff}$ scale for all the UMa stars by 900-1000 K would 
also flatten out the observed trends in [O/H] and Fe ionization state difference.  An alternative 
fix is to raise the T$_{\rm eff}$ values of the 4 coolest stars by 250 K with respect to the stars near
solar $T_{\rm eff}$.  An analogous solution to removing the [O/H] trend in the Pleiades (Schuler et al.~2004),
however, would require the cool star $T_{\rm eff}$ values to be increased by several factors of 250 K.  
We regard all these parameter variations as implausible.   

Given the totality of the evidence in Tables 2 and 4, our abundances only rule out membership 
for HD 81659.  A remaining possible curiosity is that the Fe abundances of HD 59747 are consistently 
slightly higher on a line-by-line basis compared to the similarly cool dwarfs HD 75935 and HD 173950; 
this is true for \ion{Fe}{2} as well. Such behavior, however, is not clearly seen in (e.g.,) Ca, Cr, 
or Ni.  Additional work is needed to understand how these striking anomalies noted above might also vary 
from star-to-star at a given $T_{\rm eff}$, a possibility if activity were an underlying cause (as 
suggested by, e.g. Morel \& Micela 2004 and Morel et al.~2004), within an otherwise uniform population.  
In the meantime, we note that the slightly low mean Fe abundances for the cool stars HD 28495 and HD 173950, 
the larger Fe abundances for the warmer dwarfs HD 63433 and HD 167389, and low mean Ca values for the cool 
stars HD 75395, 59747, and 173950 (all compared to the canonical UMa metallicity of [m/H]${\sim}-0.09$) that 
one might notice from Tables 2 and 4 are exactly what one expects given the overexcitation/ionization effects 
seen in, e.g., M34 (Schuler et al.~2003).  

\subsection{Li in the UMa Group} 

The Li-$T_{\rm eff}$ morphology of our UMa members (those stars in Table 7 with the exception
of HD 81659) is shown in Figure 5.  Barring the three clear members of the Li gap at 6400-6700 K, a 
notable feature seen here is the lack of statistically significant star-to-star scatter in the Li 
abundances for $T{\rm eff}{\ge}5200$ K; in this regime the spread in Li is remarkably small-- especially 
when considering the inhomogeneous data sources.  Scatter about a fitted polynomial (excluding 
Li gap stars) is only a few hundredths of a dex-- that expected from the uncertainties in 
Table 7.  
\marginpar{Fig.~5}

A second feature of note is that the 0.4-0.5 dex difference between the UMa Li abundances on the 
hot side of the F-star Li gap and the so-called Li peak at 5800-6000 K is consistent with 
that in the older Hyades and Coma clusters (e.g., Figure 3 of Jones et al.~1997 and Figures 
5 and 6 below) rather than the near-zero difference seen in the Pleiades (e.g., Figure 3 of 
King, Krishnamurthi, \& Pinsonneault 2000).  This is consistent with similar ages for UMa, 
Coma, and the Hyades as we infer from their color-magnitude diagrams.  

Below $T_{\rm eff}{\sim}5200$, Figure 5 indicates that there exists significant star-to-star 
scatter.  Figure 6 indicates the onset of this scatter occurs at similar $T_{\rm eff}$ in 
the Hyades, and the magnitude of the scatter appears similar, though the presence of 
censored data (upper limits) complicates interpretation.  What seems clear is that the 
star-to-star spread among the UMa group stars with Li detections is more similar to
the modest spread evinced by non-tidally locked binaries in the older Hyades cluster than
the large (up to a full dex) striking differences in similarly cool young (100-200 Myr) Pleiades 
and M34 cluster dwarfs (e.g., Figure 3 of Jones et al.~1997).  Our new membership information
thus verifies essentially similar conclusions of Soderblom et al.~(1993b).  
\marginpar{Fig.~6} 

The Hyades Li data in Figure 6 were analyzed in the same fashion as our UMa stars, repeating 
the analysis of \citet{Bal95} with our particular choice of $T_{\rm eff}$ scale and model 
atmosphere grids, etc.  Though UMa datapoints remain sparse, Figure 6 indicates that the 
Hyades Li abundances are lower than those in UMa for $T_{\rm eff}{\le}5400$ K.  This difference 
increases modestly if NLTE corrections \citep{Carl94} are applied.  The larger UMa Li content 
relative to cool Hyades would be further exaggerated if we were to utilize a mass coordinate, instead 
of $T_{\rm eff}$, since the masses are larger at a fixed temperature for higher metallicity like 
that characterizing the Hyades (e.g., [Fe/H]$=+0.12$ according to Cayrel, Cayrel de Strobel \& 
Campbell 1985).  Greater Li depletion in the cool Hyads is qualitatively consistent with their 
higher [Fe/H] and the well-known ``metallicity'' dependence of standard pre-main-sequence Li burning 
(e.g., Figure 3 of Chaboyer, Demarque \& Pinsonneault 1995; table 2 and \S 3.4 of Piau \& 
Turck-Chi{\`e}ze 2002); greater Li depletion would also be consistent with a very slightly older
age for the Hyades and the effects main-sequence mixing (e.g., the age dependence of Li destruction 
seen in the rotational mixing models of fixed composition in Figure 12 of Chaboyer et al.~1995).
 
While the data are very limited, comparison of the stars hotter than the F-star Li gap in
Figure 6 suggests little difference in the Hyades and UMa initial Li abundances. 
We have also noted the near-equality in the UMa and Hyades age, with an allowance that the
Hyades might be (if anything) slightly older.  It is then interesting to note that the ${\sim}0.0$ 
dex abundance difference between the cool Hyades dwarfs 5150 K and the cool UMa dwarfs at 
4850 K in Figure 6 is considerably {\it smaller} than the near ${\sim}1.0$ dex difference
predicted by the standard models in Figure 3 of Chaboyer et al.~(1995) given these entities' 
metallicity difference.  This suggests that there is a fourth parameter beyond mass, age, and 
[Fe/H] controlling relative Li depletion.  Numerous candidates abound-- helium abundance, 
detailed opacity mixtures (in particular the [$\alpha$/Fe] ratio), accretion history, initial 
angular momentum and subsequent evolution thereof.  Several of these factors are discussed and 
modeled in Piau \& Turck-Chi{\`e}ze (2002), but (unfortunately) remain observationally ill-constrained. 

Figure 7 compares our UMa Li abundances with those for Coma, which we analyzed in a
homogeneous fashion with temperatures derived as for UMa using the photometry described 
before.  The Coma Li equivalent widths were taken from cluster members in the studies 
of \citet{F01}, \citet{Sod90}, and \citet{B87}.  The UMa-Coma comparison is of particular
interest since these clusters have observationally indistinguishable age and [Fe/H]. 
Figure 7 indicates that the cool star Li abundances appear some ${\sim}0.2$ dex lower in Coma 
than the maximum abundances seen in UMa.  A similar difference is inferred from the 
relative predicted-observed Li differences for each cluster, where these differences are 
measured using the curves from Figure 3 of Chaboyer et al.~(1995) and the
two Coma stars at 5200 K and the 3 UMa stars at 4850 K.  In this case, however, the implications 
of any inter-cluster Li difference for a fourth parameter are unconvincing: the UMa-Coma comparison 
in Figure 5 may reflect a constant offset; i.e., the abundance levels on the hotside of the Li gap 
and in the G-star Li peak (5800-6000 K) may also differ at the 0.2 dex level.  Such a difference 
is suggestive of one in initial Li abundance rather than in mass-dependent Li depletion mechanisms. 
\marginpar{Fig.~7}

\subsection{Summary}

Using existing and new spectroscopy, we revisit membership and abundances for stars in 
the UMa moving group.  Comparison of the color-magnitude diagrams of UMa, Coma, and the 
Hyades using isochrones suggests that these stellar aggregates are essentially coeval, 
with the Hyades perhaps being only 100 Myr older.  This finding provides a simple explanation 
for the modest scatter of UMa stars in the chromospheric emission versus color plane compared 
to younger clusters (e.g., the Pleiades), and the indistinguishable mean chromospheric emission 
levels of UMa and Hyades members.  

Abundances from our new spectroscopy confirms non-membership for HD 81659.  Our new spectroscopy 
of field star UMa group members reveals they clearly demonstrate abundance trends suggestive of
or mimicing the effects of over-excitation/ionization that have been reported in young clusters
and very active field stars:  abundances derived from low excitation potential lines of a given
species are lower than the abundances derived from higher excitation lines; at $T_{\rm eff}{\sim}5000$ K,
Fe abundances derived from \ion{Fe}{2} lines are a factor 4 higher than abundances derived from
\ion{Fe}{1} lines; \ion{O}{1} abundances derived from the high excitation ${\lambda}7774$ triplet
show an increase with decreasing $T_{\rm eff}$.  This latter trend of rising O with declining
$T_{\rm eff}$ is strikingly muted compared to that seen in younger and more metal-rich Pleiades
stars. 

\ion{Fe}{1}-based Fe abundances for HD 28495 and 173950 are notably higher than the photometric
metallicity estimates of \citet{Nord04}.  There are several UMa members for which \citet{Nord04}
metallicities are markedly low compared to the canonical UMa abundance.  Similar oddly low
photometric estimates for cool stars are also seen in the Hyades, and the \citet{Nord04} 
photometric values yield a metallicity some 0.10-0.15 dex lower than the canonical spectroscopic
metallicity in this cluster too.    

For $T_{\rm eff}{\ge}5200$ K, UMa group member Li abundances show remarkably small dispersion 
that is compatible with the estimated errors.  As in other young clusters, however, a significant 
star-to-star scatter in Li is seen at cooler $T_{\rm eff}$ values.  Consistent redetermination of 
Hyades Li abundances indicates lower values at a given cool temperature (and corresponding mass) 
than for UMa--a difference qualitatively consistent with expectations of standard PMS burning 
given the higher Hyades [Fe/H] value.  However, the quantitative Hyades-UMa cool dwarf difference 
is considerably smaller than expected, suggesting a fourth parameter other than stellar mass, 
age, and [Fe/H] affecting the relative Li depletion in cool young-to-intermediate age open cluster 
dwarfs.  Cool dwarfs in UMa and Coma, which have observationally indistinguishable age and [Fe/H], 
show only a modest ${\le}0.2$ dex, if any, difference that could be due to an initial abundance difference.  
 
Uniquely identifying this fourth parameter may be observationally challenging:  the initial 
angular momentum distribution and details of angular momentum loss are both folded into the 
present day stellar rotation distributions (which may additionally be convolved with projection 
effects); observable signatures of accretion will quickly be lost by convective dilution in 
sufficiently low-mass stars; and stellar He abundances are notoriously difficult to determine 
(particularly in low mass stars).  Differences in detailed opacity mixtures are the most amenable 
to observational discrimination, but consistent abundance analyses of numerous elements in large 
samples of open cluster stars do not yet exist.  While this lack of important observational data 
seems easy to remedy, the abundance results presented here and the recent work indicating (presumably 
spurious) $T_{\rm eff}$- and/or age- and/or activity-dependent trends in cool dwarf abundances 
(Cayrel et al.~1985; King et al.~2000; Schuler et al.~2003,2004; Yong et al.~2004; Morel \& Micela 
2004) suggests the needed delineation of genuine cluster-to-cluster abundance differences is not
necessarily straightforward.  Indeed, an interesting future question is how any such effects influence 
the measured Li abundances in different clusters themselves. 

\acknowledgments
The author gratefully acknowledges support for this work from NSF awards AST-0086576 and AST-0239518, and 
a generous grant from the Charles Curry Foundation to Clemson University.  We also thank Dr.~Aaron 
Steinhauer for kindly providing his {\sf{LIFIND}} code, and Ms. Abigail Daane and Mr. Roggie Boone
for their assistance at the McDonald 2.7-m and KPNO 4-m telescopes.

\begin{figure}
\plotone{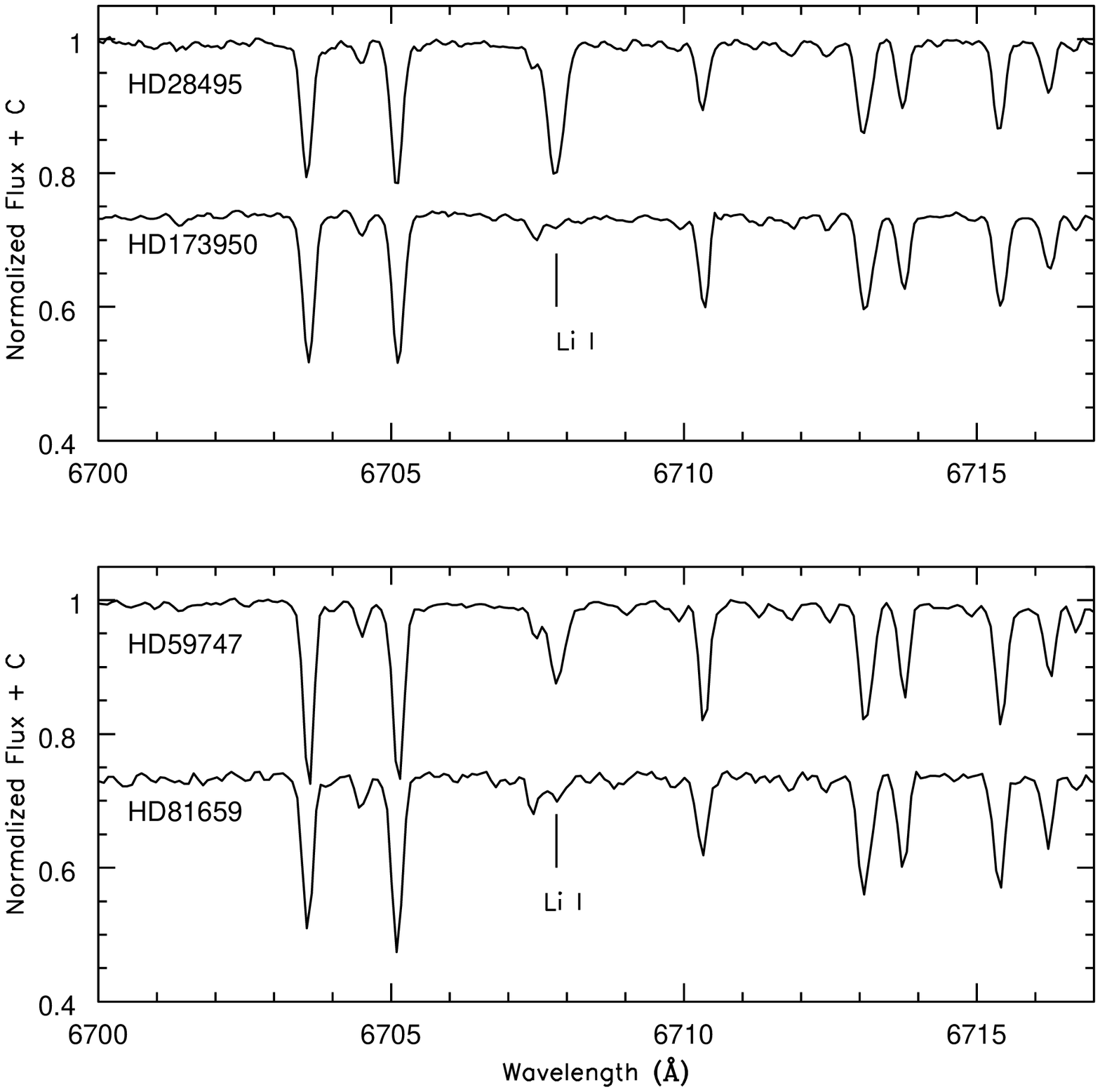}
\caption[]{Sample spectra of our additional UMa candidates obtained at the McDonald Observatory 2.7m (top panel) and 
Kitt Peak National Observatory 4m (bottom panel).}   
\end{figure}

\begin{figure}
\plotone{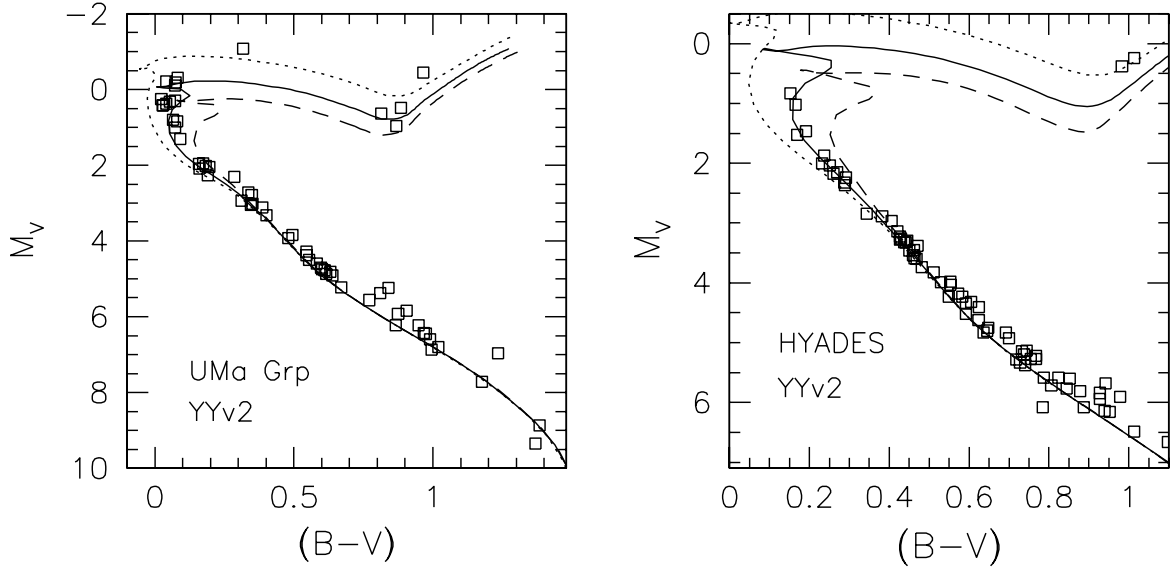}
\caption[]{(Left) The Hipparcos-based UMa group color-magnitude diagram is shown with the 400, 600, and 
800 Myr Yale-Yonsei [Fe/H]$=-0.08$ isochrones constructed with the Lejeune et al.~(1998) color-temperature
relation. (Right) The Hipparocs based Hyades color-magnitude diagram is shown with the 500, 700, and 900 Myr 
Yale-Yonsei [Fe/H]$=+0.13$ isochrones constructed with the same color-temperature relation.  }  
\end{figure}

\begin{figure}
\plotone{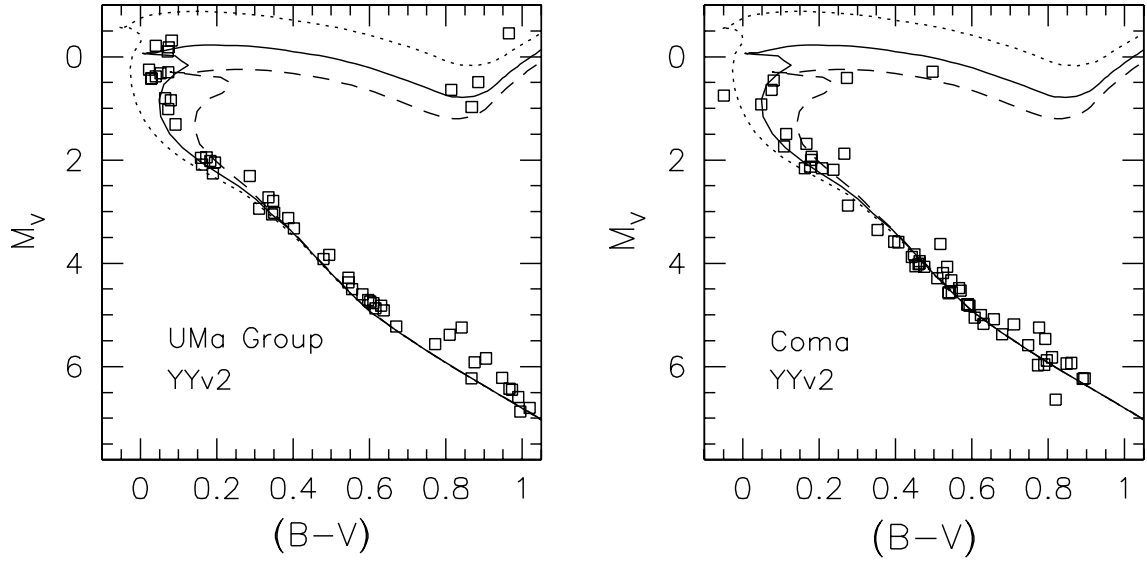}
\caption[]{(Left) The UMa color magnitude diagram from Figure 1 is shown again.  (Right) The Coma 
color-magnitude diagram, assuming $(m-M)=4.54$ and $E(B-V)=0.00$, is plotted with the same 
400, 600, and 800 Myr [Fe/H]$=-0.08$ isochrones as for UMa.  } 
\end{figure} 

\begin{figure}
\plotone{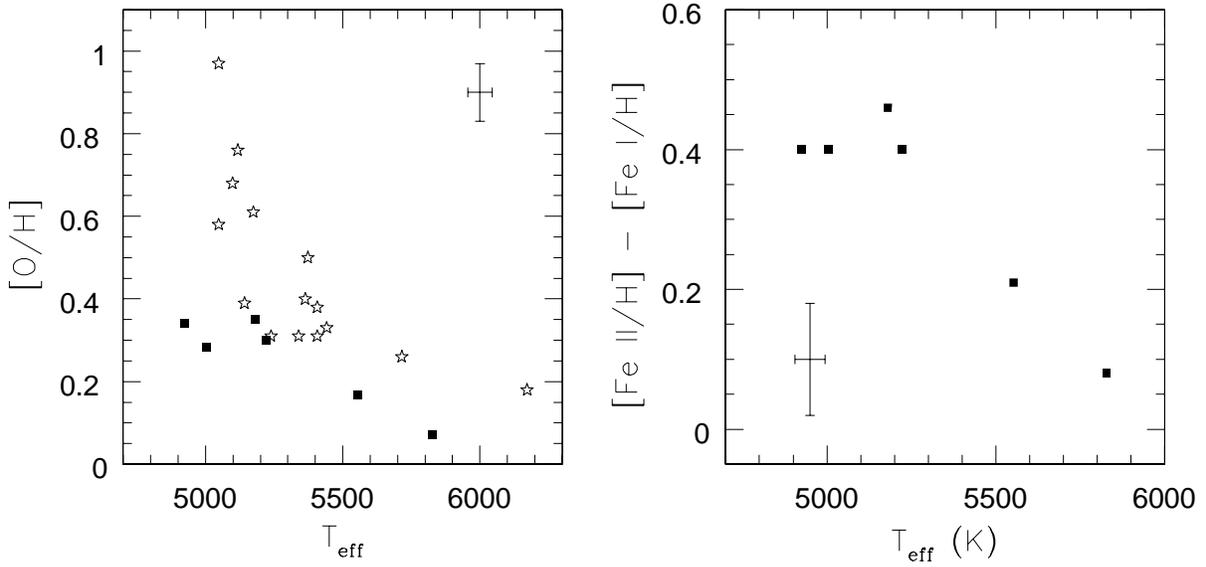}
\caption[]{Left-- The ${\lambda}7774$ \ion{O}{1}-based [O/H] values from our own spectroscopy are plotted
versus $T_{\rm eff}$ for UMa group objects (filled squares) and Pleiades dwarfs from Schuler et 
al.~(2004; open stars); a typical error bar is shown in the upper right.  Right-- The difference between 
[Fe/H] determined from \ion{Fe}{2} and \ion{Fe}{1} lines based on our own spectroscopy of UMa group 
objects.  A typical error bar is shown in the bottom left.} 
\end{figure}

\begin{figure}
\plotone{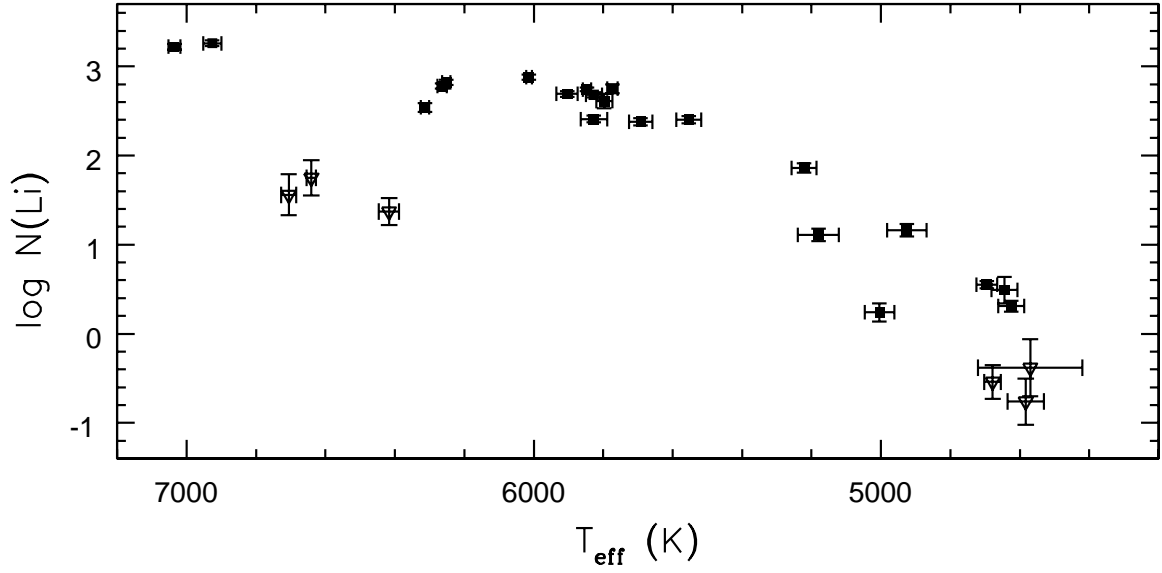}
\caption[]{LTE Li abundance is plotted versus $T_{\rm eff}$ for our UMa objects.  Upper limits
are shown as inverted open triangles.}
\end{figure}

\begin{figure}
\plotone{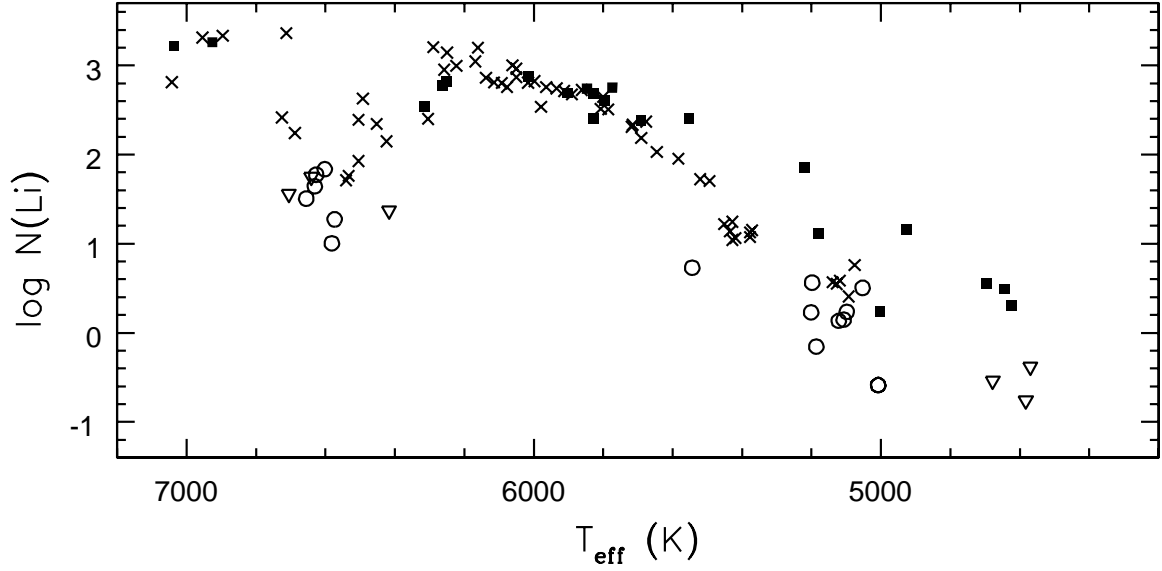}
\caption[]{LTE Li abundances for the Hyades (crosses and open circles; the latter designating upper limits)
and our UMa objects (symbols the same as in Figure 3) are shown versus $T_{\rm eff}$.} 
\end{figure}

\begin{figure}
\plotone{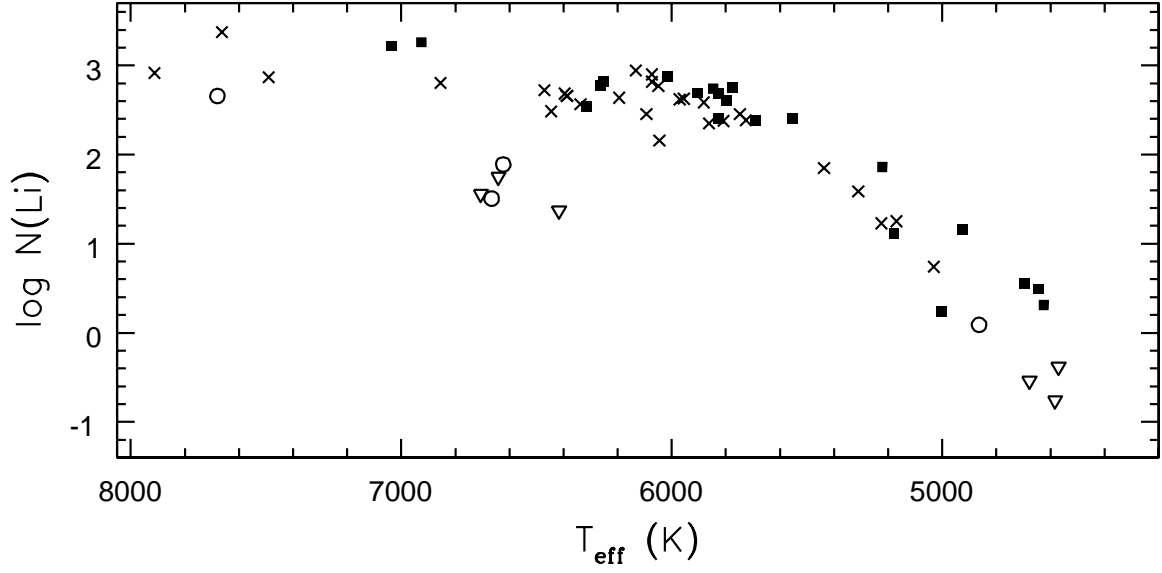}
\caption[]{LTE Li abundances for Coma Berenices (crosses and open circles denote detections and upper limits)
and our UMa objects (with symbols as in Figure 3) are shown versus $T_{\rm eff}$.}
\end{figure}

\begin{deluxetable}{lrrrrrrrrrr}
\tablecolumns{11}
\tablewidth{0pc}
\tablenum{1}
\tablecaption{UMa Group \ion{Fe}{1} Line Data} 
\tablehead{
\colhead{$\lambda$} & \colhead{$\chi$} & \colhead{log $gf$} & \colhead{HD28495} & \colhead{HD59747} & \colhead{HD63433} & \colhead{HD75935} & \colhead{HD81659} & \colhead{HD167389} & \colhead{HD173950} & \colhead{Sun}\\
\colhead{\AA} & \colhead{eV} & \colhead{ } & \colhead{m\AA} & \colhead{m\AA} & \colhead{m\AA} & \colhead{m\AA} & \colhead{m\AA} & \colhead{m\AA} &
\colhead{m\AA} & \colhead{m\AA}
}
\startdata
6703.58 & 2.76 & -3.13 & 49.9 & 65.3 & \nodata & 54.2 & 53.7 & 35.5    & 56.0 & 40.3 \\       
6713.75 & 4.79 & -1.52 & 24.3 & 31.8 & 25.6    & 28.7 & 33.1 & 19.6    & 27.2 & 23.0 \\ 
6725.36 & 4.10 & -2.30 & 22.4 & 30.4 & 23.9    & 26.1 & 27.0 & 16.9    & 24.7 & 19.6 \\
6726.67 & 4.61 & -1.12 & 55.0 & 69.4 & 57.4    & 62.3 & 58.5 & 48.1    & 57.3 & 50.5 \\ 
6739.52 & 1.56 & -4.98 & 21.6 & 33.7 & 14.7    & 26.4 & 23.8 & 10.6    & 26.9 & 12.8 \\ 
6745.98 & 4.07 & -2.74 & 8.4  & 13.5 & 6.7     & 10.8 & 12.0 & 6.2     & 9.4  & 7.5  \\  
6746.98 & 2.61 & -4.35 & 6.6  & 11.6 & 6.7     & 7.5  & 9.0  & \nodata & 9.2  & 4.3  \\  
\enddata
\end{deluxetable}

\begin{deluxetable}{lrrrrrrrr}
\tablecolumns{9}
\tablewidth{0pc}
\tablenum{2}
\tablecaption{UMa Group Parameters and Fe Abundance Data} 
\tablehead{
\colhead{ } & \colhead{HD28495} & \colhead{HD59747} & \colhead{HD63433} & \colhead{HD75935} & \colhead{HD81659} & \colhead{HD167389} & \colhead{HD173950} & \colhead{Sun}\\
}
\startdata
Parameters & & & & & & & & \\
$T_{\rm eff}$ & 5222 & 4925  & 5553  & 5180  & 5467  & 5827  & 5004  & 5770 \\ 
log $g$       & 4.62 & 4.65  & 4.57  & 4.63  & 4.58  & 4.52  & 4.65  & 4.44 \\  
${\xi}$       & 1.14 & 1.04  & 1.25  & 1.12  & 1.22  & 1.36  & 1.07  & 1.38 \\  
$[{\rm m/H}]$ & -0.2 & -0.10 & -0.10 & -0.10 & +0.10 & -0.10 & -0.20 & +0.00 \\ 
\cline{1-9}
Abundances & & & & & & & & \\
6703.58    & -0.18 & +0.01 & \nodata & -0.10 & +0.04 & -0.03   & -0.16 & 7.59 \\  
6713.75    & -0.16 & +0.02 & -0.02   & -0.05 & +0.11 & -0.05   & -0.09 & 7.59 \\ 
6725.36    & -0.17 & +0.00 & +0.01   & -0.07 & +0.04 & -0.04   & -0.15 & 7.62 \\ 
6726.67    & -0.15 & +0.06 & +0.02   & -0.02 & +0.02 & -0.01   & -0.14 & 7.56 \\ 
6739.52    & -0.21 & +0.08 & -0.13   & -0.11 & +0.05 & -0.03   & -0.21 & 7.57 \\ 
6745.98    & -0.20 & +0.02 & -0.16   & -0.07 & +0.07 & -0.05   & -0.18 & 7.55 \\ 
6746.98    & -0.19 & -0.01 & +0.04   & -0.13 & +0.12 & \nodata & -0.12 & 7.47 \\ 
\cline{1-9} 
Results & & & & & & & & \\
$[{\rm Fe/H}]$ & -0.18 & +0.03 & -0.04 & -0.08 & +0.06 & -0.04 & -0.15 & 7.564 \\  
std dev & 0.020 & 0.033 & 0.084 & 0.038 & 0.037 & 0.015 & 0.039 & 0.048 \\ 
\enddata
\end{deluxetable}

\begin{deluxetable}{lrrrrrrr}
\tablecolumns{8}
\tablewidth{0pc}
\tablenum{3}
\tablecaption{Supplemental Line Data} 
\tablehead{
\colhead{Species} & \colhead{$\lambda$} & \colhead{$\chi$} & \colhead{log $gf$} & \colhead{Sun} & \colhead{HD75935} & \colhead{HD59747} & \colhead{HD173950} \\
\colhead{ } & \colhead{\AA} & \colhead{eV} & \colhead{ } & \colhead{EW(m{\AA})} & \colhead{EW(m{\AA})} & \colhead{EW(m{\AA})} & \colhead{EW(m{\AA})} 
}
\startdata
\ion{Ca}{1} & 6417.69\tablenotemark{a} & 4.44 & -0.75 & 15.0  & 18.0  & 22.6  & 19.3 \\  
            & 6449.82 & 2.52 & -0.62 & 109.0 & 134.0 & 156.9 & 145.7 \\
            & 6455.61 & 2.52 & -1.50 &  58.9 & 75.0  & 93.4  & 82.5 \\
            & 6464.68 & 2.52 & -2.53 &  15.2 & 22.6  & 34.9  & 27.4 \\
            & 6499.65 & 2.52 & -1.00 &  89.7 & 111.0 & 126.5 & 121.7 \\
\ion{Cr}{1} & 6330.10 & 0.94 & -2.99 &  30.8 &  51.2 &  71.8 & 59.3 \\
            & 6661.08 & 4.19 & -0.24 &  13.0 & \nodata &  28.1 & 23.6 \\
            & 6729.75 & 4.39 & -0.66 &   3.6 &   5.4 &   6.7 &  5.8 \\ 
\ion{Fe}{1} & 6498.95 & 0.96 & -4.70 &  50.0 &  69.2 &  83.5 & 75.1 \\
            & 6608.04 & 2.28 & -4.02 &  19.7 &  31.3 &  40.1 & 33.9 \\
            & 6609.12 & 2.56 & -2.67 &  71.1 &  87.7 &  99.9 & 90.8 \\
\ion{Fe}{2} & 6416.93 & 3.89 & -2.86 &  45.2 &  39.5 &  34.6 & 32.0 \\
\ion{Ni}{1} & 6327.60 & 1.68 & -3.23 &  40.6 &  54.0 &  60.5 & 49.2 \\
            & 6378.26 & 4.15 & -1.00 &  34.3 &  37.6 &  38.3 & 34.6 \\
            & 6414.59 & 4.15 & -1.29 &  20.0 &  23.9 &  21.2 & 20.6 \\
            & 6482.81 & 1.93 & -2.97 &  45.1 &  52.8 &  57.2 & \nodata \\
            & 6532.88 & 1.93 & -3.47 &  18.4 &  27.6 &  31.3 & 23.5 \\
            & 6598.61 & 4.23 & -1.02 &  27.5 &  31.8 &  28.5 & 27.5 \\
            & 6635.14 & 4.42 & -0.87 &  27.3 &  29.7 &  28.0 & 24.8 \\
            & 6767.78 & 1.83 & -1.89 &  83.5 &  94.3 & 105.4 & 95.6 \\ 
\tablenotetext{a}{Appears clean in the Sun, but blended to the red in the cool UMa stars.  This is 
accounted for in the measured equivalent widths by reflecting fit to the blue side of the line profile.}
\enddata
\end{deluxetable}                 

\begin{deluxetable}{lrrrrr}
\tablecolumns{6}
\tablewidth{0pc}
\tablenum{4}
\tablecaption{Supplemental Abundance Results} 
\tablehead{
\colhead{Species} & \colhead{$\lambda$} & \colhead{Sun} & \colhead{HD75935} & \colhead{HD59747} & \colhead{HD173950} \\
\colhead{ } & \colhead{\AA} & \colhead{log $N$} & \colhead{[X/H]} & \colhead{[X/H]} & \colhead{[X/H]} 
}
\startdata
\ion{Ca}{1} & 6417.69 & 6.63 & -0.10 & -0.07 & -0.13 \\  
            & 6449.82 & 6.31 & -0.24 & -0.31 & -0.33 \\  
            & 6455.61 & 6.44 & -0.19 & -0.16 & -0.24 \\  
            & 6464.68 & 6.59 & -0.20 & -0.15 & -0.23 \\  
            & 6499.65 & 6.41 & -0.19 & -0.28 & -0.27 \\  
mean [Ca/H] &         &      & -0.18 & -0.19 & -0.24 \\  
std dev     &         &      & 0.051 & 0.099 & 0.073 \\
\ion{Cr}{1} & 6330.10 & 5.78 & -0.27 & -0.20 & -0.23 \\  
            & 6661.08 & 5.71 & \nodata & -0.01 & -0.08 \\      
            & 6729.75 & 5.71 & -0.08 & -0.08 & -0.12 \\  
mean [Cr/H] &         &      & -0.18 & -0.10 & -0.14 \\  
std dev     &         &      & \nodata & 0.096 & 0.077 \\
\ion{Fe}{1} & 6498.95 & 7.54 & -0.20 & -0.10 & -0.23 \\  
            & 6608.04 & 7.57 & -0.15 & -0.07 & -0.19 \\  
            & 6609.12 & 7.49 & -0.10 & -0.03 & -0.16 \\  
mean [Fe/H]\tablenotemark{a} &         &      & -0.10 & -0.00 & -0.16 \\ 
std dev     &         &      & 0.052 & 0.055 & 0.042 \\     
\ion{Fe}{2} & 6416.93 & 7.75 & +0.40 & +0.59 & +0.37 \\  
\ion{Ni}{1} & 6327.60 & 6.37 & -0.05 & +0.01 & -0.22 \\  
            & 6378.26 & 6.44 & -0.04 & +0.01 & -0.10 \\  
            & 6414.59 & 6.40 & 0.00  & -0.03 & -0.09 \\  
            & 6482.81 & 6.43 & -0.13 & -0.10 & \nodata \\  
            & 6532.88 & 6.36 & -0.02 & -0.03 & -0.23 \\  
            & 6598.61 & 6.38 & 0.00  & -0.02 & -0.09 \\  
            & 6635.14 & 6.41 & -0.03 & -0.02 & -0.13 \\ 
            & 6767.78 & 5.92 & -0.11 & -0.03 & -0.18 \\  
mean [Ni/H] &         &      & -0.05 & -0.03 & -0.15 \\  
std dev     &         &      & 0.048 & 0.034 & 0.061 \\
\tablenotetext{a}{Includes the results for Fe features in Table2.}
\enddata
\end{deluxetable}

\begin{deluxetable}{lrrr} 
\tablecolumns{4} 
\tablewidth{0pc}
\tablenum{5}
\tablecaption{UMa LTE \ion{O}{1} Abundances}
\tablehead{
\colhead{Quantity} & \colhead{${\lambda}7772$} & \colhead{${\lambda}7774$} & \colhead{${\lambda}7775$}
}
\startdata
log $gf$             & +0.333 & +0.186 & -0.035 \\ 
Solar EW(m{\AA})     & 73.7   & 63.8   & 50.1   \\ 
log $N$(O)$_{\odot}$ & 8.94   & 8.93   & 8.91   \\  
HD28495 EW           & 52.2   & 42.7   & 32.4   \\  
$[{\rm O/H}]$              & +0.31  & +0.29  & +0.30  \\  
HD59747 EW           & 30.9   & 28.0   & 18.5   \\  
$[{\rm O/H}]$                & +0.32  & +0.39  & +0.32  \\  
HD63433 EW           & 65.8   & 58.1   & 45.5   \\  
$[{\rm O/H}]$                & +0.14  & +0.17  & +0.19  \\  
HD75935 EW           & 49.0   & 42.3   & 33.1   \\  
$[{\rm O/H}]$                & +0.33  & +0.36  & +0.39  \\ 
HD167389 EW          & 82.0   & 71.5   & 55.6   \\  
$[{\rm O/H}]$                & +0.07  & +0.07  & +0.07  \\  
HD173950 EW          & 33.8   & 28.0   & 23.3   \\   
$[{\rm O/H}]$                & +0.25  & +0.25  & +0.35  \\   
\enddata
\end{deluxetable} 

\begin{deluxetable}{lrrrr} 
\tablecolumns{5} 
\tablewidth{0pc}
\tablenum{6}
\tablecaption{UMa \ion{Fe}{2} Abundances}
\tablehead{
\colhead{Quantity} & \colhead{${\lambda}6149.25$} & \colhead{${\lambda}6247.56$} & \colhead{${\lambda}6416.93$} & \colhead{${\lambda}6456.39$}
}
\startdata
log $gf$             & -2.72  & -2.31 & -2.86 & -2.08 \\ 
Solar EW(m{\AA})     & 38.2   & 56.6  & 45.2  & 67.7  \\ 
log $N$(Fe)$_{\odot}$ & 7.47  & 7.44  & 7.75  & 7.43  \\  
HD28495 EW           & 28.7   & 46.3  & 34.7  & 59.1  \\   
$[{\rm Fe/H}]$       & +0.18  & +0.23 & +0.19 & +0.28 \\  
HD59747 EW           & 22.2   & 32.7  & 34.6  & 43.1  \\  
$[{\rm Fe/H}]$       & +0.35  & +0.30 & +0.59 & +0.35 \\  
HD63433 EW           & 40.7   & 55.1  & 41.6  & 67.6  \\  
$[{\rm Fe/H}]$       & +0.23  & +0.16 & +0.10 & +0.19 \\ 
HD75935 EW           & 29.8   & 49.1  & 39.5  & 57.8  \\  
$[{\rm Fe/H}]$       & +0.29  & +0.38 & +0.40 & +0.35 \\ 
HD167389 EW          & 40.2   & 62.0  & 43.8  & 72.2  \\  
$[{\rm Fe/H}]$       & +0.03  & +0.10 & -0.04 & +0.08 \\  
HD173950 EW          & 20.6   & 34.2  & 32.0  & 45.3  \\  
$[{\rm Fe/H}]$       & +0.16  & +0.19 & +0.37 & +0.25 \\ 
\enddata
\end{deluxetable}

\begin{deluxetable}{lrrrrrcccccc}
\tablecolumns{12}
\tablewidth{0pc}
\tablenum{7}
\tablecaption{UMa Group Li Data} 
\tablehead{
\colhead{Star} & \colhead{[Fe/H]} & \colhead{[Fe/H]} & \colhead{$(B-V)$} & \colhead{$T_{\rm eff}$} & \colhead{EW(Li)} & \colhead{Li Ref} & 
\colhead{log N(Li)} & \colhead{log $R'_{\rm HK}$} & \colhead{Ref} & 
\colhead{$v$ sin $i$} & \colhead{Ref} \\
\colhead{HD} & \colhead{Phot} & \colhead{Spect} & \colhead{ } & \colhead{K} & \colhead{m{\AA}} & \colhead{ } & \colhead{LTE} & \colhead{ } & \colhead{ } & \colhead{km/s} & \colhead{ } 
}
\startdata
11131   & -0.27 & -0.10 & 0.638 & 5692${\pm}34$  & 67.5                      & 1,2   &                 & $-4.50$ & 3,4,5 & $3.5$ & 6 \\ 
        &       &       &       &                & 78.0\tablenotemark{a}     & 7,8,9 & 2.38${\pm}0.04$ &         &       &       &   \\
13959A  &       &       & 0.995 & 4570${\pm}150$ & ${\le}6$\tablenotemark{b} & 1     & ${\le}-0.38{\pm}0.32$    & $-4.35$ & 6     & $5.0$ & 6 \\
26923   & -0.08 & +0.08 & 0.582 & 5904${\pm}31$  & 84.3                  & 1,2,10,11 & 2.69${\pm}0.03$ & $-4.50$ & 2,3,6 & $4.0$ & 2,6,11,12 \\
38393   & +0.01 & -0.07 & 0.495 & 6252${\pm}12$ & 66.5                  & 13,14 &                 & $-4.77$ & 3 & $8.7$ & 15  \\
        &       &       &       &               & 58.5\tablenotemark{a} & 1,16  & $2.82{\pm}0.03$ &         &   &       &     \\ 
38392   &       &       & 0.954 & 4678${\pm}24$ & ${\le}3$              & 1,16  & ${\le}-0.54{\pm}0.19$     & $-4.48$ & 5,17,18 & $1.8$ & 17 \\
39587   & -0.06 & -0.04 & 0.597 & 5847${\pm}12$ & 100.3                 & 14,19 &                 & $-4.41$ & 3,6     & $7.5$ & 6, 20 \\
        &       &       &       &               & 102.8\tablenotemark{a} & 1,8  & 2.74${\pm}0.02$ &         &         &       & \\ 
72905   & -0.20 & -0.05 & 0.616 & 5774${\pm}15$ & 108.5                  & 2         &                 & $-4.37$ & 2,3,6 & $9.3$ & 2,6,21 \\
        &       &       &       &               & 116.9\tablenotemark{a} & 1,8,22,23 & 2.75${\pm}0.05$ &         &       &       & \\
109011  &       &       & 0.948 & 4695${\pm}30$ & 23.4                   & 1,2       & 0.55${\pm}0.04$ & $-4.37$ & 2,3,6 & $5.4$ & 2,6 \\
109647  &       &       & 0.967 & 4644${\pm}37$ & 25.0                   & 1         & 0.49${\pm}0.15$ & $-4.45$\tablenotemark{c} & 3,18  & $2.3$ & 6   \\ 
109799  & -0.24 & -0.08 & 0.336 & 6926${\pm}26$ & 56.0\tablenotemark{a}  & 8         & 3.26${\pm}0.03$ & $-4.36$\tablenotemark{d} & 24    & $0.0$ & 25 \\
110463  &       &       & 0.974 & 4625${\pm}37$ & 18.0                   & 1         & 0.31${\pm}0.06$ & $-4.43$\tablenotemark{c} & 3,6,26 & $2.1$ & 6 \\
111456  & -0.18 &       & 0.480 & 6313${\pm}12$ & 36\tablenotemark{a}    & 1         & 2.54${\pm}0.05$ & $-4.38$\tablenotemark{d} & 6,24 & $35$ & 6 \\   
115043  & -0.19 & -0.03 & 0.610 & 5797${\pm}23$ & 77\tablenotemark{e}    & 1         &                 & $-4.45$ & 3,6 & $7.5$ & 6\\
        &       &       &       &               & 101\tablenotemark{a}   & 22        & 2.61${\pm}0.08$ & & & & \\  
125451A & -0.02 & -0.02 & 0.402 & 6641${\pm}13$ & ${\le}3.9$\tablenotemark{a} & 8    & ${\le}1.750{\pm}0.20$ & $-4.37$\tablenotemark{d} & 24 & $43$  & 15,28 \\
129798A & +0.00 &       & 0.387 & 6706${\pm}22$ & ${\le}2.4$\tablenotemark{a} & 29   & ${\le}1.56{\pm}0.23$  & \nodata                  &    & $43$  & 30  \\
141003B &       &       & 0.99  & 4583${\pm}52$ & ${\le}2$               & 1         & ${\le}-0.76{\pm}0.26$ & $-4.38$\tablenotemark{c} & 18 & $3.3$ & 6 \\
147584  & -0.08 & -0.19 & 0.554 & 6014${\pm}8$  & 101                    & 31        &                       & $-4.56$                  & 5  & $2.2$ & 32 \\
        &       &       &       &               & 93                     & 33        & $2.88{\pm}0.03$       & & & & \\
165185  & -0.21 & -0.06 & 0.602 & 5827${\pm}23$ & 92.7                   & 2         &                       & $-4.45$                  & 2,3 & $7.2$ & 2 \\
        &       &       &       &               & 93.0                   & 1         & $2.68{\pm}0.03$       &   & & & \\ 
180777  &       &       & 0.311 & 7035${\pm}17$ & 45.4\tablenotemark{a}  & 8         & $3.22{\pm}0.03$       & -4.34\tablenotemark{d}   & 24 & 63  & 24 \\ 
184960  & -0.32 & -0.14 & 0.492 & 6264${\pm}13$ & 57.0                   & 1,34      &                       & $-5.07$\tablenotemark{c} & 18 & ${\le}7$ & 6 \\
        &       &       &       &               & 61.6\tablenotemark{a}  & 8         & $2.78{\pm}0.03$       &   & & & \\
211575  & +0.09 &       & 0.455 & 6417${\pm}29$ & ${\le}2.7$             & 35        & ${\le}1.37{\pm}0.15$  & $-4.71$ & 36 & $18$ & 37 \\ 
\cline{1-10}
New Data &  & & & & & & & & \\ 
28495   & -0.41 & -0.18 & 0.772 & 5222${\pm}36$ & 71.5                   & 38        & $1.86{\pm}0.05$       & $-4.39$ & 3  &  &   \\ 
59747   & -0.14 & -0.00 & 0.867 & 4925${\pm}57$ & 41.0                   & 39        & $1.16{\pm}0.07$       & $-4.44$ & 3  &  &   \\
63433   & -0.10 & -0.04 & 0.676 & 5553${\pm}36$ & 99.6                   & 39        & $2.40{\pm}0.04$       & $-4.42$ & 3  &  &   \\ 
75935   &       & -0.10 & 0.785 & 5180${\pm}59$ & 21.9                   & 39        & $1.11{\pm}0.07$       & $-4.44$ & 3  &  &   \\
81659   & +0.13 & +0.06 & 0.700 & 5467${\pm}49$ & 17.2                   & 39        & $1.32{\pm}0.09$       & $-4.57$ & 3  &  &   \\
167389  & -0.11 & -0.04 & 0.602 & 5827${\pm}38$ & 58.8                   & 38        & $2.41{\pm}0.04$       & $-4.74$ & 3  &  &   \\
173950  & -0.43 & -0.16 & 0.841 & 5004${\pm}43$ & 5.8                    & 38        & $0.24{\pm}0.10$       & $-4.46$ & 3  &  &   \\
\enddata
\tablerefs{ 
(1) Soderblom et al.~(1993); (2) Gaidos, Henry \& Henry (2000); (3) King et al.~(2003); 
(4) Tinney et al.~(2002); (5) Henry et al.~(1996); (6) Soderblom \& Mayor (1993a);  
(7) Pallavicini, Randich, \& Giampapa (1992); (8) Boesgaard, Budge \& Burck (1988); 
(9) Boesgaard \& Tripicco (1987); (10) Favata et al.~(1995); (11) Randich et al.~(1999);
(12) L{\`e}bre et al.~(1999); (13) Soderblom, King \& Henry (1998); (14) Chen et al.~(2001); 
(15) Soderblom, Pendleton \& Pallavicini (1989) ; (16) Pallavicini, Cerruti-Sola \& Duncan (1987); 
(17) Soderblom \& Mayor (1993b); (18) Soderblom \& Clements (1987); (19) Lambert, Heath \& Edvardsson (1991); 
(20) Strassmeier et al.~(1990); (21) Fekel (1997); (22) Montes et al.~(2001); 
(23) Wichmann, Schmitt \& Hubrig (2003); (24) Simon \& Landsman (1991); (25) Uesugi \& Fukuda (1970); 
(26) Strassmeier et al.~(2000); (28) de Medeiros \& Mayor (1999); (29) Russell (1995)  
(30) Royer et al.~(2002); (31) Rebolo et al.~(1986); (32) Saar \& Osten (1997); 
(33) Soderblom (1985); (34) Balachandran (1990); (35) Deliyannis et al.~(1998);
(36) Soderblom, Duncan, \& Johnson (1991); (37) Wolff \& Simon (1997); (38) McDonald 2.7-m; (39) KPNO 4-m 
}

\tablenotetext{a}{Equivalent width measurement contains a contribution from the nearby 6707.4{\AA} \ion{Fe}{1}$+$CN blending features}
\tablenotetext{b}{Review of the resolved photometry in Fabricius \& Makarov (2000) suggests the close components
of HD 13959AB have near equal brightness at 6700 {\AA}.  The original equivalent width upper limit has thus been
doubled to account for continuum dilution.}  
\tablenotetext{c}{The log R'$_{hk}$ index from (18) has been transformed to log R'$_{HK}$ using their relations.}
\tablenotetext{d}{The log R'$_{1335}$ index from (24) has been transformed to log R'$_{hk}$ and then to log R'$_{HK}$ using the relations in (18).}  
\tablenotetext{e}{The 77 m{\AA} equivalent width from (1) differs substantially from the Fe-corrected equivalent width of 96 m{\AA} from (22).} 

\end{deluxetable}
\end{document}